# Synthesis of nanographite by laser induced thermal expansion of $H_2SO_4$ intercalated graphite lattice

G. Carotenuto[1], A. Longo[1], L. Nicolais[1]

[1]Institute for Polymers, Composites and Biomaterials - National Research Council.

Mostra d'oltremare Pad 20, Viale Kennedy 54, 80125 Naples. Italy

S. De Nicola[2],

[2]SPIN - National Research Council,

Complesso Universitario Monte Sant'Angelo, Via Cinthia, 80126 Naples. Italy.

E. Pugliese[3], M. Ciofini[3], M. Locatelli[3], A. Lapucci[3], R. Meucci[3],

[3]National Institute of Optics - National Research Council.

Largo E. Fermi 6, 50125 Florence, Italy

**Abstract.**

We have experimentally investigated the expansion mechanism induced by laser irradiation of $H_2SO_4$ intercalated graphite. Intercalated graphite materials irradiated by laser allows to study reactions occurring in confined chemical conditions. We find that the laser-assisted reaction process is characterized by a threshold temperature of about 140°C followed by a fast rate of heating which produces a large amount of overheated gases and causes a violent expansion of the graphite crystal. We investigate the morphological changes undergone by irradiated graphite filaments using a combination of thermal and visible imaging technique which allows for the quantitative determination of temporal evolution of the thermal field during the heating stage and exfoliation of the grain.

# 1. Introduction

Graphene is a new and very promising multifunctional material that can be easily obtained from graphite flakes by various chemical methods. A chemical route makes use of three main steps: (1) graphite intercalation by sulfuric acid after mild oxidation by nitric acid or other strong oxidizer such as potassioum chlorate or potassium permangamate; (2) thermal expansion of sulfuric acid intercalated graphite to obtain expanded graphite (EG) and (3) sonoacustic exfoliation of the expanded graphite filaments[1-3].

The most important stage is the formation of the thermally expanded graphite, though the nature of the final product ( graphene monolayers, few-layer graphene or graphite nanoplatelets) depends also on the intercalation order.

Thermal expansion of sulfuric acid intercalated graphite consists in separating the non intercalated nanoregions in the graphite lattice, thus reducing the adhesion between the generated nanocrystals. The presence of graphite intercalated compounds (CIGs) is required for a successful exfoliation process. In fact, heating the CIG induces the reaction the carbon and sulfuric with formation of a mixture of overheated gases acid and the pressure generated from the evolved gases causes a significant expansion of the material along the crystallographic axis.

An important issue is to investigate the mechanism involved in the above described preparative scheme. We explore a laser assisted method for thermal exfoliation of $H_2SO_4$ intercalated graphite under ambient atmosphere. The lamellar structure of graphite provide the environment that allows the carbon/ $H_2SO_4$ reaction to take place in confined conditions under laser irradiation [4-7]. The reaction takes place according to the following scheme [8]

$$2 H_2SO_4 + C = 2 H_2O + CO_2 + 2 SO_2 \qquad (1)$$

This reaction leads to the formation of gaseous molecules $H_2O$, $SO_2$, and $CO_2$. The process can be understood as a redistribution of the bonds between the atoms of the intercalated reactants with the laser irradiation supplying the activation energy of the reaction and the strong heating of the gaseous species followed by the violent expansion of the crystal [9-12].

The above reaction occurs in confined conditions, i.e, when the sulfuric acid is dissolved into the graphite crystals. However, chemical reactions occurring in confined conditions behaves quite differently compared to the same under ordinary conditions. In same cases, confinement facilitate chemical reactions that, otherwise, don't occur for kinetic reasons even if they are thermodynamically allowed. Indeed, graphite *per se*, does not react with sulfuric acid. A well known example is the dehydration reaction of sugar by sulfuric acid which stops at carbon stage without formation of $H_2O$, $CO_2$, and $SO_2$ gaseous mixture [13-14].

On the other hand, the confinement of $H_2SO_4$ molecules into the graphite lattice enhances the kinetic of carbon-$H_2SO_4$ reaction. In fact, the graphite lattice potential provides tight quantum confinement of the $H_2SO_4$ molecules and it is expected that, upon decreasing the volume of the available space, the energy of the intercalated reactants increases. Therefore, the activation energy is lowered and the reaction proceeds at enhanced rate. Fig. 1 depicts qualitatively the change in energy due to the redistribution of the bonds between the reactant species.

Since the reaction is an endothermic process, heat is required to allow the chemical transformation and the reaction rate is essentially limited by the heat flow into the reactive carbon/ $H_2SO_4$ system.

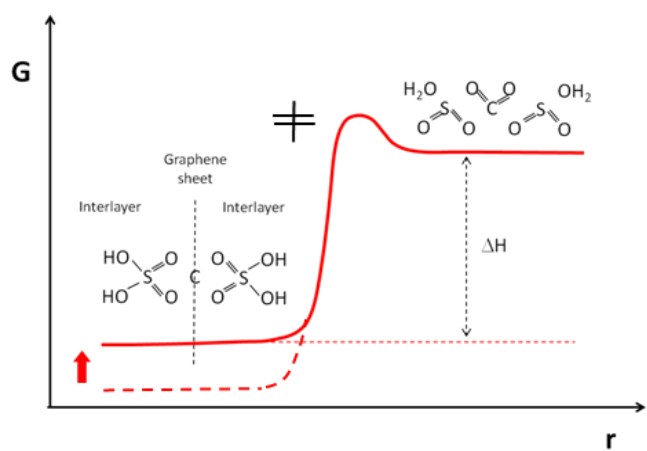

**Fig. 1** - Reaction coordinate diagram for the carbon/$H_2SO_4$/ chemical reaction.

Here, we describe an infrared (IR) laser assisted method for thermal exfoliation of $H_2SO_4$ intercalated graphite and experimentally investigate the thermal behavior of the graphite intercalated compounds during crystal expansion process. We show that the chemical reaction in the intercalated graphite can be directly observed using a thermal imaging technique which allows for the quantitative determination of temporal evolution of the thermal field during the heating stage and exfoliation of the grain.

## 2. Experimental part

The laser induced thermal exfoliation process was studied on single intercalated graphite grains by heating them using a medium-power infrared laser source. In particular, a 100 W Yitterbium laser (YLR-100-SM-AC by IPG Photonics) with a wavelength of 1.07 µm was used, and the laser output power was increased by steps of 0.2 W. (see Fig. 2). Due to the small size of the irradiated grains ( about 1 mm), an high resolution microscope, equipped with a Pyroelectric camera (Spiricon Pyrocam III) and an high-speed visible video camera, were used. The system was specifically designed to allow for simultaneously recording both visible and infrared images and to correlate the evolution of the grain expansion with the temporal behavior of the thermal field during the laser irradiation. Fig. 2 ( right side) shows the temperature calibration curve. The process was monitored in the mid-infrared region (8-12 µm) in order to separate the relatively weak thermal emission of the grains from the large amount of scattered laser radiation. IR videos (see supporting information) were recorded up to a maximum rate of 24 frames/s by using a Ge close-up lens system. Visible videos (see supporting information) were recorded at high rate (250 frames/s) by a camera equipped with macro lens.

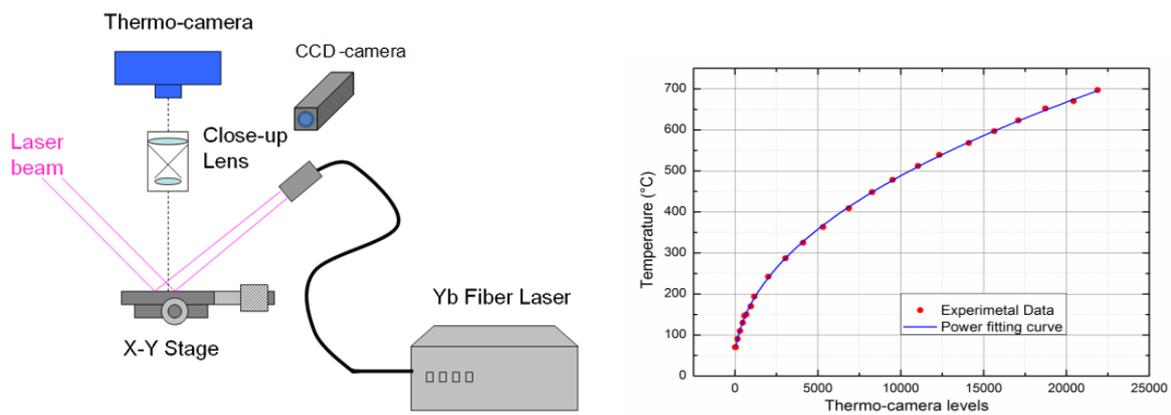

**Fig. 2 -** Experimental setup (left-side picture) and temperature calibration curve (right-side picture).

The $H_2SO_4$-intercalated graphite samples were provided by Faima (Italy) and were used in the "as provided" form. The temporal behavior of the thermal field was investigated by image analysis of sequences of frames recorded by the infrared video camera. The grainl was stuck at bottom surface by silicon grease to avoid movement of the sample during the laser irradiation. Both the grease and the plate were perfectly transmitting at the exciting laser wavelength.

The morphological characterization of the intercalated graphite samples before and after laser treatment was performed by scanning electron microscopy (SEM, FEI Quanta 200 FEG equipped with an Oxford Inca Energy System 250).

## 3. Results and discussion

Figure 3a shows a sequence of thermal images recorded at different time during the grain explosion process. The grain is exposed to a laser power density of 16 W/cm$^2$. The temperature field changes significantly during the laser irradiation and the thermal images show clearly that the temperature across the grain rises very quickly. The bright red area in the IR images corresponds to the high temperature hot spot across the grain volume. The frame sequence also shows the external region surrounding the hot grain, which becomes colder as the grain expansion process proceeds. This region corresponds to the thermal field of the gas mixture $H_2O$, $CO_2$, and $SO_2$ which are removed during the explosion of the grain.

Fig 3b shows in more detail the temporal evolution of the temperature across the irradiated grain. At the beginning of the laser induced heating process the temperature of the grain rises slowly until a critical threshold is reached and the chemical reaction is activated. The experimental data clearly show that once the reaction starts there is a quick rise in temperature from the threshold value (about 140 °C) to a value of about 350 °C at 2.67s, after which the temperature starts decreasing slightly eventually reaching a thermal equilibrium.

To improve the accuracy in the determination of the temperature threshold, several measurements were performed on different grain under the same irradiation condition. Fig. 4 show an histogram of the threshold temperature measurements. The measurements results were collected from laser irradiation tests carried out on 24 grains.

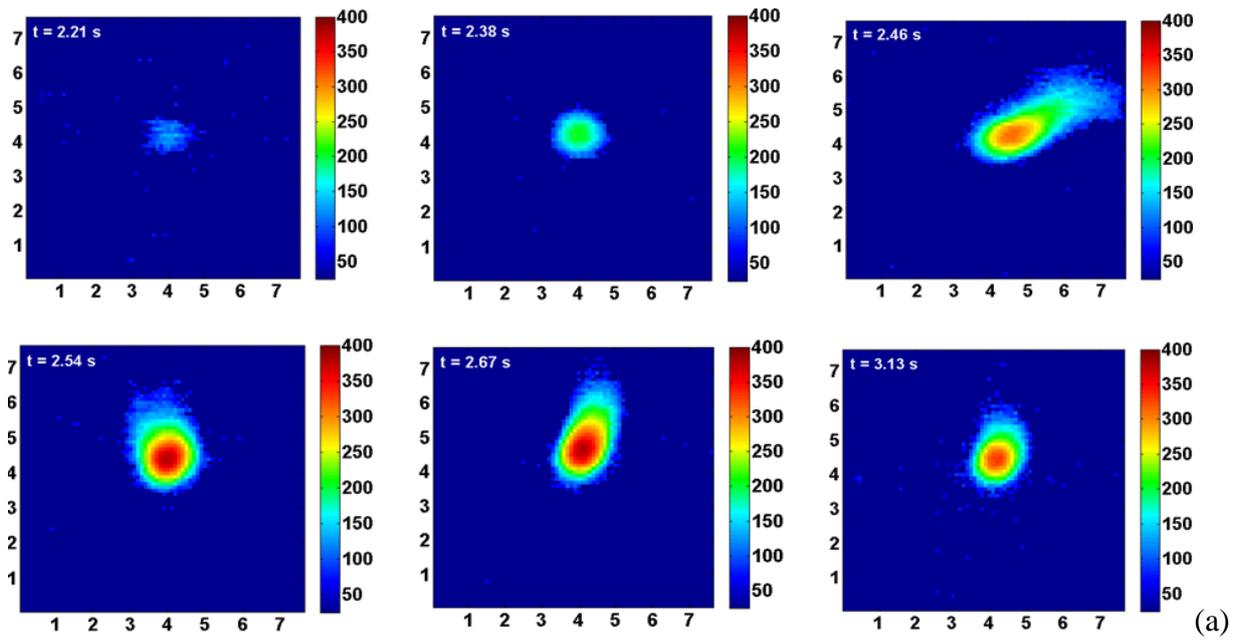

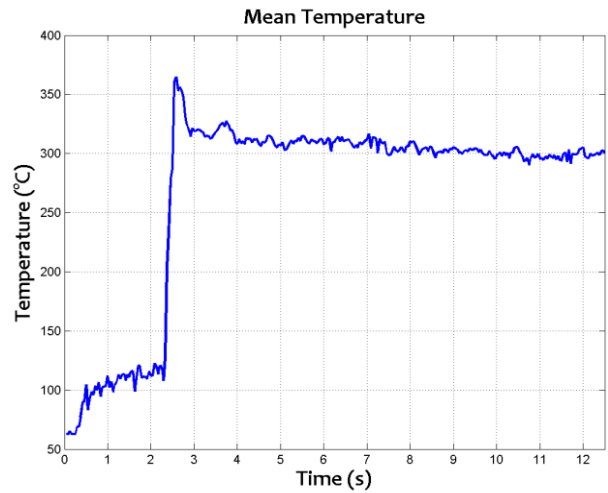

**Fig. 3** – (a) Series of thermal images recorded at different times during the irradiation process (7.5 × 7.5mm field of view). The grain is exposed to a laser power density of 16 W/cm$^2$; (b) plot of the temporal evolution of the laser-irradiated grain temperature.

The temperature threshold depends on the grain size and therefore on its ability to dissipate heat during the thermal treatment. Such an observation is of a certain technological relevance since it shows that all techniques that can be used to heat the intercalated graphite crystals produce completely equivalent materials.

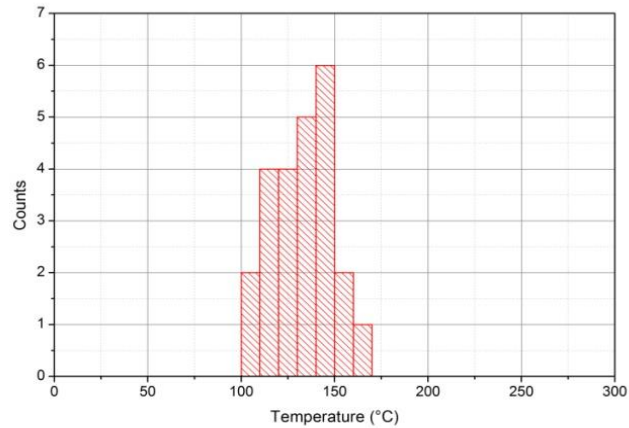

**Fig. 4 –** Hystogram of measurements of threshold temperature: the statistical analysis was based on a 24 graphite flakes population.

The laser-induced expansion process is also studied by analyzing the frame sequence of the irradiated grain, recorded in the visible range. Fig. 5 shows the sequence of images of the graphite filament exposed to laser irradiation. The sequence of these frames, recorded at the same time of in the infrared ((shown in fig. 3a), confirms that the expansion does not proceeds gradually. The distinctive feature of the process is a threshold time after which the overheated $H_2O$, $SO_2$, and $CO_2$ gases are produced quite instantaneously (within about 0.2 sec) and completely expand the crystal in a burst ( compare sequences 3-5).

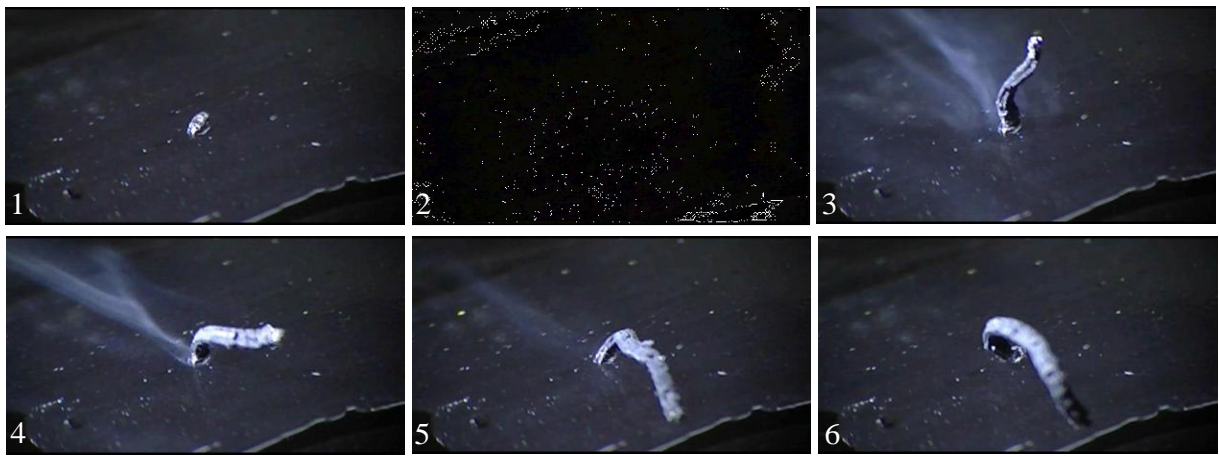

**Fig. 5 -** Micrographs of the laser-induced expansion process. The frame sequence from 1 to 6 corresponds to the sequence of thermal images shown in fig. 3a

During the filament expansion the gaseous mixture is released in the flake almost isotropically, while the expanded crystal undergoes various kind of violent deformations such as bending and torsion [15] (see supporting video information). The time evolution of the length of the extruded flake ($L_n - L_0$, where $L_n$ is length at frame n) is shown in fig. 6. A length increase by 650% was obtained by analyzing the whole temporal sequence of the frames by an image processing software (ImageJ).

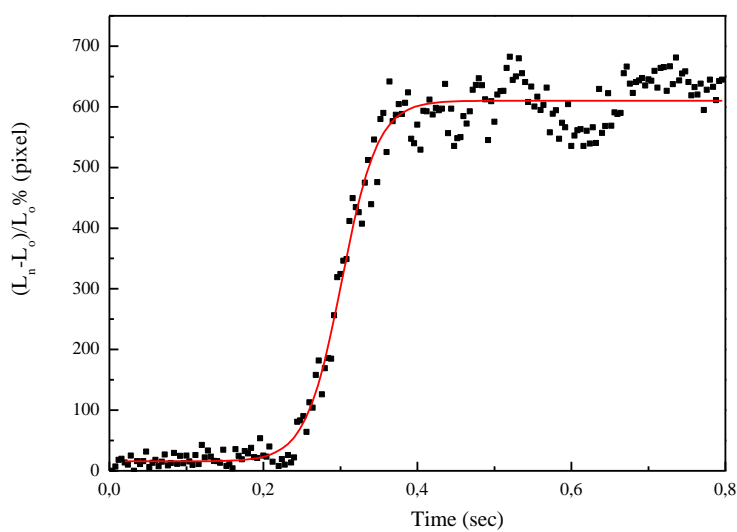

**Fig. 6** Plot of the temporal evolution of the grain length

The frame sequence shows that the lamellar crystal expansion starts at the top of the crystal and it proceeds along the c-crystallographic axis. This behavior can be clearly understood: the endothermic reaction requires heat which is provided by absorption of the laser radiation impinging on the lamella top. Heat transfer proceeds from the top to the lamella at the bottom of the crystal. Diffusion of heat inside the crystals results in the redistribution of the Carbon/$H_2SO_4$ bonds with formation of the expanding gaseous molecules $CO_2$, $H_2O$, and $SO_2$.

The energy supplied by laser irradiation promotes reaction of $H_2SO_4$ molecules contained inside the graphite crystal. We observe that the crystal lattice expands violently but it does not explode. After the release of the gas in the environment the whole crystal structure has a volume slightly smaller compared to that at the instant of the gas expulsion, indicating that the nanographite lattice did not change significantly.

Scanning electron microscopy (SEM) have been used to analyze the different morphologies of samples before and after the laser induced thermal expansion. Figures 7A and 7B show SEM micrographs of a sample before the laser irradiation. It can be clearly seen a disc-like platelet and its typical layered structure. Typically, the platelets had thickness of 50 μm and the size of 500 μm.

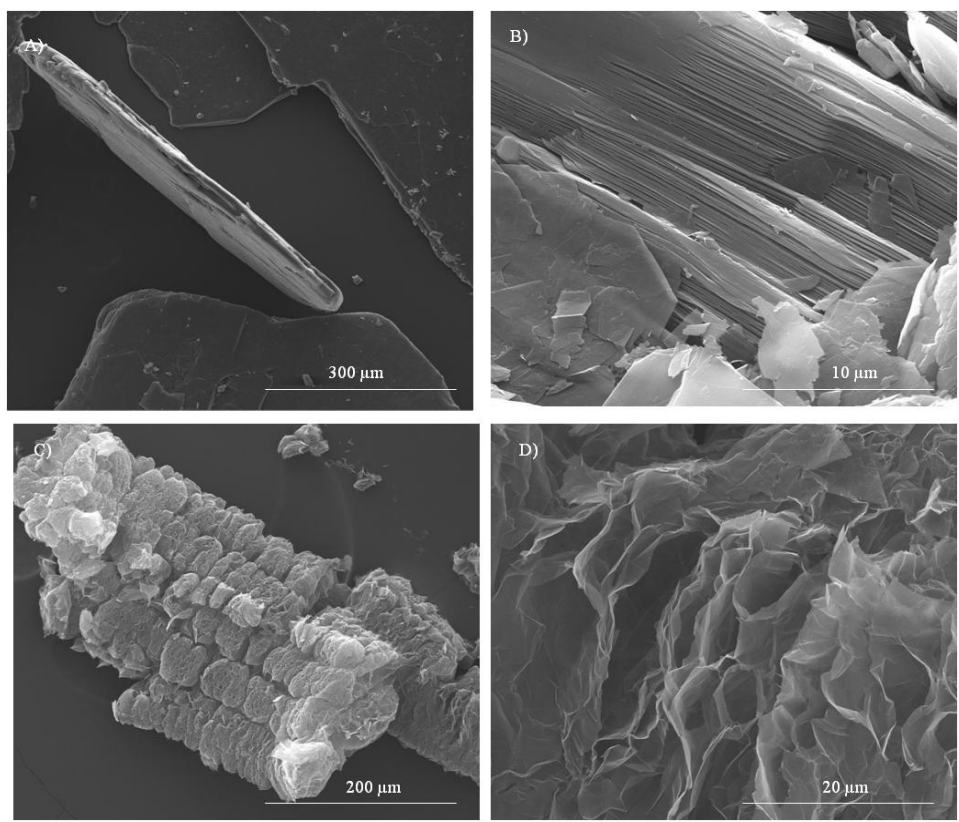

**Figure 7**. (A) (B) SEM micrographs at different magnification of GICs; (C), (D) SEM of expanded graphite after the laser irradiation

After the laser irradiation it is noticeable the unidirectional expansion induced by the laser, resulting in a sort of worm-like filament (see figure 7C), i.e., a long, porous, deformed cylinder like structure. Figure 7D shows the expanded structure at higher magnification to make more visible the full expansion between each layer and the presence of open channels or cavities. Physically the number, the size and the distribution of such cavities along the worm like structure depend on the intercalate compounds [16]. In addition, the cavities have an ellipsoid cross sections [17].

An over-heating of the molecules in the gas mixture is expected to increase the porosity content of the expanded crystals. Such effect could be profitably used in the preparation of graphite nano-platelets because high porosity graphite structure can be detached more easily by the sonication treatment.

## 4. Conclusion

We have experimentally investigated the expansion mechanism induced by laser irradiation of $H_2SO_4$ intercalated graphite. We have shown that the chemical reaction in the intercalated graphite can be directly observed using a thermal imaging technique which allows for the quantitative determination of temporal evolution of the thermal field during the heating stage and exfoliation of the grain. Intercalated graphite materials provides an interesting scenario for studying confined chemical reactions. Laser irradiation promotes reaction of $H_2SO_4$ molecules contained inside the graphite crystal by heating the irradiated samples. The capability of the monitoring system to follow the various stage of the heating process allows to obtain useful information regarding the thermal expansion dynamics. In particular, the mixture of gases $H_2O$, $SO_2$, and $CO_2$ produced during the thermal expansion stage of the $H_2SO_4$-intercalated graphite flakes, is mainly heated by the large amount of thermal energy released by the carbon-$H_2SO_4$ chemical reaction. The laser-assisted reaction process is characterized by a threshold temperature followed by a fast rate of heating which produces a large amount of overheated gases and causes a violent expansion of the graphite crystal.